\documentclass{elsart}
\journal{New Astronomy}
\usepackage{natbib}

\usepackage{epsfig}

\usepackage{amssymb}
\def\be{\begin{equation}}
\def\ee{\end{equation}}
\def\bea{\begin{eqnarray}}
\def\eea{\end{eqnarray}}
\def\biposh#1#2#3{ A^{#1}_{#2|#3}}
\def\tildebiposhmat#1#2#3{ {\bf\tilde{A}}^{#1}_{#2|#3}}
\def\biposhmat#1#2#3{ {\bf A}^{#1}_{#2|#3}}
\begin{document}

\begin{frontmatter}



 \title{Measuring Statistical Isotropy of CMB Anisotropy}


\author{{\bf Tarun Souradeep}$^1$, Amir Hajian$^{2,3}$ and Soumen Basak$^1$}
\address{$^1$Inter-University Centre for Astronomy and Astrophysics
(IUCAA),\\ Post Bag 4, Ganeshkhind, Pune 411~007, India.\\  E-mail:
tarun@iucaa.ernet.in;soumen@iucaa.ernet.in \\$^2$ Department of Physics,
Jadwin Hall, Princeton University, Princeton, NJ 08542. \\$^3$ Department
of Astrophysical Sciences, Peyton Hall, Princeton
University,Princeton, NJ 08544.\\ E-mail:ahajian@princeton.edu}
\begin{abstract}
% Text of abstractve 

The statistical expectation values of the temperature fluctuations and
polarization of cosmic microwave background (CMB) are assumed to be
preserved under rotations of the sky. We investigate the statistical
isotropy (SI) of the CMB maps recently measured by the Wilkinson
Microwave Anisotropy Probe (WMAP) using the bipolar spherical harmonic
formalism proposed in Hajian \& Souradeep 2003 for CMB temperature
anisotropy and extended to CMB polarization in Basak, Hajian \&
Souradeep 2006. The {\em Bipolar Power Spectrum (BiPS)} had been
measured for the full sky CMB anisotropy maps of the first year WMAP
data and now for the recently released three years of WMAP data.  We
also introduce and measure directional sensitive {\em reduced Bipolar
coefficients} on the three year WMAP ILC map. Consistent with our
published results from first year WMAP data we have no evidence for
violation of statistical isotropy on large angular scales. Preliminary
analysis of the recently released first WMAP polarization maps,
however, indicate significant violation of SI even when the foreground
contaminated regions are masked out. Further work is required to
confirm a possible cosmic origin and rule out the (more likely) origin
in observational artifact such as foreground residuals at high
galactic latitude.

\end{abstract}


\end{frontmatter}

% main text
\section{Introduction}

In standard cosmology, CMB anisotropy signal is expected to be
statistically isotropic, i.e., statistical expectation values of the
temperature fluctuations $\Delta T(\hat q)$ are preserved under
rotations of the sky. In particular, the angular correlation function
$C(\hat{q},\, \hat{q}^\prime)\equiv\langle\Delta T(\hat q)\Delta
T(\hat q^\prime)\rangle$ is rotationally invariant for Gaussian
fields. In spherical harmonic space, where $\Delta T(\hat q)=
\sum_{lm}a_{lm} Y_{lm}(\hat q)$ the condition of {\em statistical
isotropy} (SI) translates to a diagonal $\langle a_{lm} a^*_{l^\prime
m^\prime}\rangle=C_{l} \delta_{ll^\prime}\delta_{mm^\prime}$ where
$C_l$, is the widely used angular power spectrum of the CMB
anisotropy. Statistical isotropy of the CMB sky is essential for $C_l$
to be a complete description of (Gaussian) CMB anisotropy and, hence,
an adequate measure for carrying out cosmological parameter estimation
of `standard' (SI) model. Hence, it is crucial to be able to determine
from the observed CMB sky whether it is a realization of a
statistically isotropic process, or not. The detection of statistical
isotropy (SI) violations in the CMB signal can have exciting and
far-reaching implications for cosmology and the cosmological
principle.  For example, a generic consequence of cosmic topology is
the breaking of statistical isotropy in characteristic patterns
determined by the photon geodesic structure of the manifold as probed
by the CMB photons traveling to us from the surface of last scattering
over a distance comparable to the cosmic horizon,
$R_H$~\cite{bps98,bps00a,bps00b}.  Mildly anisotropic cosmological
models predict charateristic patterns hidden in the CMB sky. On the
other hand, SI violation could also arise from foreground
contamination, or, artifacts of observational and analysis techniques.

The CMB measurements of the {\it Wilkinson Microwave Anisotropy Probe}
 ({\it WMAP}) are consistent with predictions of the concordance
 $\Lambda$CDM model with (nearly) scale-invariant and adiabatic
 fluctuations expected to have been generated during an inflationary
 epoch in the early universe~\cite{hin_wmap03, kogut_wmap03,
 sper_wmap03, page_wmap03,
 peiris_wmap03,sper_wmap06,hin_wmap06,pag_wmap06}. After the first
 year of {\it WMAP} data, the SI of the CMB anisotropy ({\it i.e.}
 rotational invariance of n-point correlations) has attracted
 considerable attention.  Tantalizing evidence of SI breakdown
 (albeit, in very different guises) has mounted in the {\it WMAP}
 first year sky maps, using a variety of different
 statistics~\cite{erik04a, Copi:2003kt, Schwarz:2004gk, Hansen:2004vq,
 angelwmap, Land:2004bs,Land:2005ad, Land:2005dq,
 Land:2005jq,Land:2005cg, Bielewicz:2004en,
 Bielewicz:2005zu,Copi:2005ff, Copi:2006tu, Naselsky:2004gm,
 Prunet:2004zy,Gluck:2005td, Stannard:2004yp, Bernui:2005pz,
 Bernui:2006ft, Freeman:2005nx, Chen:2005ev,Wiaux:2006zh}. The
 three-year WMAP maps are consistent with the first-year maps up to a
 small quadrupole difference. The two additional years of data and the
 improvements in analysis has not significantly altered the low
 multipole structures in the maps~\cite{hin_wmap06}. Hence,
 `anomalies' are expected to persist at the same modest level of
 significance but are now less likely to be artifacts of noise,
 uncorrected systematics, or the analysis of the first year data.  The
 cosmic significance of these `anomalies', however, remains debatable
 because of the aposteriori statistics employed to ferret them out of
 the data. {\em More importantly, what is missing is a common, well
 defined, mathematical language to quantify SI (as distinct from non
 Gaussianity) and the ability to ascribe statistical significance to
 the anomalies unambiguously.} We employ a well defined, mathematical
 language of Bipolar harmonic decomposition of the underlying
 correlation to quantify SI that can ascribe statistical significance
 to the anomalies unambiguously.

The observed CMB sky is a single realization of the underlying
correlation, hence the detection of SI violation or correlation
patterns pose a great observational challenge. For statistically
isotropic CMB sky, the correlation function
\begin{equation}
C(\hat{n}_1,\hat{n}_2)\equiv C(\hat{n}_1\cdot\hat{n}_2) =
\frac{1}{8\pi^2}\int d{\mathcal R} C({\mathcal R}\hat{n}_1,\,
{\mathcal R}\hat{n}_2),
\label{avg_cth}
\end{equation}
where ${\mathcal R}\hat{n}$ denotes the direction obtained under the
action of a rotation ${\mathcal R}$ on $\hat{n}$, and $d{\mathcal R}$
is a volume element of the three-dimensional rotation group.  The
invariance of the underlying statistics under rotation allows the
estimation of $C(\hat{n}\cdot\hat{n}')$ using the average of the
temperature product $\widetilde{\Delta T}(\hat n) \widetilde{\Delta
T}(\hat n')$ between all pairs of pixels with the angular separation
$\theta$.  In the absence of statistical isotropy,
$C(\hat{n},\hat{n}')$ is estimated by a single product
$\widetilde{\Delta T}(\hat n)\widetilde{\Delta T}(\hat n')$ and hence
is poorly determined from a single realization. Although it is not
possible to estimate each element of the full correlation function
$C(\hat{n},\hat{n}')$, some measures of statistical anisotropy of the
CMB map can be estimated through suitably weighted angular averages of
$\widetilde{\Delta T}(\hat n)\widetilde{\Delta T}(\hat n')$. The
angular averaging procedure should be such that the measure involves
averaging over sufficient number of independent `measurements', but
should ensure that the averaging does not erase all the signature of
statistical anisotropy.  We proposed the Bipolar Power spectrum (BiPS)
$\kappa_\ell$ ($\ell=1,2,3, \ldots$) of the CMB map as a statistical
tool of detecting and measuring departure from
SI~\cite{us_apjl,us_pascos} and reviewed in this article in
sec.~\ref{bips}. The non-zero value of the BiPS spectrum imply the
break down of statistical isotropy
\begin{equation} 
{\mathrm {\Huge STATISTICAL\,\,\,\, ISOTROPY}} \,\,\,\,\,\,\, 
\Longrightarrow \,\,\,\,\,\,\, 
\kappa_\ell\,=\,0 \,\,\,\,\,\,\, \forall \ell \ne 0.
\label{bipsstate}
\end{equation}
The BiPS is sensitive to structures and patterns in the underlying
total two-point correlation function \cite{us_apjl, us_pascos}.  The
BiPS is particularly sensitive to real space correlation patterns
(preferred directions, etc.) on characteristic angular scales. In
harmonic space, the BiPS at multipole $\ell$ sums power in
off-diagonal elements of the covariance matrix, $\langle a_{lm}
a_{l'm'}\rangle$, in the same way that the `angular momentum' addition
of states $l m$, $l' m'$ have non-zero overlap with a state with
angular momentum $|l-l'|<\ell<l+l'$. Signatures, like $a_{lm}$ and
$a_{l+n m}$ being correlated over a significant range $l$ are ideal
targets for BiPS. These are typical of SI violation due to cosmic
topology and the predicted BiPS in these models have a strong spectral
signature in the bipolar multipole $\ell$ space~\cite{us_prl}.  The
orientation independence of BiPS is an advantage for constraining
patterns (preferred directions) with unspecified orientation in the
CMB sky such as that arising due to cosmic topology or, anisotropic
cosmology~\cite{ghos06}.

The results of WMAP are a milestone in CMB anisotropy measurements
since it combines high angular resolution, high sensitivity, with
`full' sky coverage allowed by a space mission. The frequency coverage
allows for WMAP CMB sky maps to be foreground cleaned up to $l\sim
100$~\cite{angelwmap,sah06}. The CMB anisotropy map based on the WMAP
data are ideal for testing statistical isotropy.  Measurement of
the BiPS on CMB anisotropy maps based on the first year WMAP data are
consistent with statistical isotropy~\cite{us_apjl2,us_apj,us_jgrg}.  BIPS
analysis on the recent WMAP-3yr data are consistent with SI and
further, indicate that BiPS of the three years maps show an
improvement in SI -- the deviations are smaller and
fewer~\cite{haj_sour06}. 

First `full-sky' CMB polarization maps have been recently delivered by
WMAP~\cite{pag_wmap06}. The statistical isotropy of the CMB
polarization maps are an independent probe of statistical isotropy.
Since CMB polarization is generated on at the surface of last
scattering, violations of statistical isotropy are pristine cosmic
signatures and more difficult to attribute to the local universe.  The
bipolar power spectrum has been defined and implemented for CMB
polarization and show great promise~\cite{bas06}.

\section{The Bipolar representation of SI violation} 
\label{bips}

Although it is not possible to estimate each element of the full
correlation function $C(\hat{n}_1,\hat{n}_2)$, some measures of
statistical isotropy of the CMB map can be estimated through suitably
weighted angular averages of $\Delta T(\hat{n}_1) \Delta
T(\hat{n}_2)$. The angular averaging procedure should be such that the
measure involves averaging over sufficient number of independent
measurements to reduce the cosmic variance, but should ensure that the
averaging does not erase all the signature of statistical
anisotropy. The Bipolar power spectrum (BiPS) is a measure of
statistical isotropy proposed by us in ref.~\cite{us_apjl}
 \begin{equation}\label{kl}
 \kappa^{\ell}\,=\, (2l+1)^2 \int d\Omega_{n_1}\int d\Omega_{n_2} \,
 [\frac{1}{8\pi^2}\int d{\mathcal R} \chi^{\ell}({\mathcal R})\,
 C({\mathcal R}\hat{n}_1,\, {\mathcal R}\hat{n}_2)]^2
 \end{equation}
that also has another important desirable property of being
independent of the overall orientation of the sky.  In the above
expression, $ C({\mathcal R}\hat{n}_1,\, {\mathcal R}\hat{n}_2)$ is
the two point correlation at ${\mathcal R}\hat{n}_1\,$ and $ {\mathcal
R}\hat{n}_2$ which are the coordinates of the two pixels $\hat{n}_1\,$
and $\hat{n}_2$ after rotating the coordinate system by element
${\mathcal R}$ of the rotation group. $\chi^{\ell}$ is the trace of
the finite rotation matrix in the $\ell M$-representation
$\chi^{\ell}({\mathcal R})\,=\,\sum_{M=-\ell}^{\ell}
D_{MM}^{\ell}({\mathcal R})$ which is called the {\it characteristic
function}, or the character of the irreducible representation of rank
$\ell$.  For a statistically isotropic model,  $ C(\hat{n}_1,\,
\hat{n}_2)$ is invariant under rotation, and therefore $ C({\mathcal
R}\hat{n}_1,\, {\mathcal R}\hat{n}_2)\,=\,C(\hat{n}_1,\, \hat{n}_2)$
and then the orthonormality of $\chi^{\ell}(\omega)$, implies the
condition for SI, $ \kappa^{\ell} \, = \, \kappa^0 \delta_{\ell 0}$
stated in eq.(\ref{bipsstate}).

The harmonic space representation BiPS provides complementary
understanding and allows for rapid numerical computation.  Two point
correlations of CMB anisotropy, $C(\hat{n}_1,\, \hat{n}_2)$, are
functions on $S^2 \times S^2$, and hence can be expanded as \be
\label{bipolar} C(\hat{n}_1,\, \hat{n}_2)\, =\, \sum_{l_1,l_2,\ell,M}
\biposh{}{\ell M}{l_1 l_2} Y^{l_1l_2}_{\ell M}(\hat{n}_1,\,
\hat{n}_2)\,. \ee Here $\biposh{}{\ell M}{l_1 l_2}$ are the Bipolar
Spherical harmonic (BipoSH) coefficients of the expansion and
$Y^{l_1l_2}_{\ell M}(\hat{n}_1,\, \hat{n}_2)$ are bipolar spherical
harmonics. Bipolar spherical harmonics form an orthonormal basis on
$S^2 \times S^2$ and transform in the same manner as the spherical
harmonic function with $\ell,\, M$ with respect to rotations
\cite{Var}. Consequently, inverse-transform of $C(\hat{n}_1,\,
\hat{n}_2)$ in eq.~(\ref{bipolar}) to obtain the BipoSH coefficients
of expansion is unambiguous.

Most importantly, the Bipolar Spherical Harmonic (BipoSH)
coefficients, $\biposh{}{\ell M}{l_1 l_2}$, are linear combinations of
{\em off-diagonal elements} of the harmonic space covariance matrix,
\be
\label{ALMvsalm} \biposh{}{\ell M}{l_1 l_2} \,=\, \sum_{m_1m_2}
\langle a_{l_1m_1}a^{*}_{l_2 m_2}\rangle (-1)^{m_2} {\mathcal C}^{\ell
M}_{l_1m_1l_2 -m_2} \ee where ${\mathcal C}_{l_1m_1l_2m_2}^{\ell M}$
are Clebsch-Gordan coefficients. This clearly shows that
$\biposh{}{\ell M}{l_1 l_2}$ completely represent the information of
the covariance matrix.  

Statistical isotropy implies that the covariance matrix is diagonal, $
\langle a_{lm}a^{*}_{l' m'}\rangle = C_{l}\,\, \delta_{ll^\prime}
\delta_{mm'}$ and hence the angular power spectra carry all
information of the field. When statistical isotropy holds BipoSH
coefficients, $\biposh{}{\ell M}{ll'}$, are zero except those with
$\ell=0, M=0$ which are equal to the angular power spectra up to a
$(-1)^l (2l+1)^{1/2}$ factor.  Therefore to test a CMB map for
statistical isotropy, one should compute the BipoSH coefficients for
the maps and look for nonzero BipoSH coefficients. {\em Statistically
significant deviations of BipoSH coefficient of map from zero would
establish violation of statistical isotropy.}
 
It is impossible to measure all $A^{\ell M}_{l_1 l_2}$ individually
from the single CMB sky map because of cosmic variance since they form
an equivalent representation of a general two point correlation
function. There are several ways of combining BipoSH coefficients that
serve to highlight different aspects of SI violations. Further, one
can exploit the freedom to smooth the map in real space by appropriate
symmetric kernels $W(\hat n\cdot\hat n')$ to target specific regions
of the multipole space by the corresponding window function $W_l$.

\subsection{Bipolar Power spectrum (BiPS)}
 
Among the several possible combinations of BipoSH coefficients, the
Bipolar Power Spectrum (BiPS) has proved to be a useful tool with
interesting features \cite{us_apjl}. BiPS of CMB anisotropy is defined
as a convenient contraction of the BipoSH coefficients \be
\label{kappal} \kappa_\ell \,=\, \sum_{l,l',M} W_l W_{l'}\left|\biposh{}{\ell
M}{ll'}\right|^2 \geq 0 \ee where $W_l$ is the window function that
corresponds to  smoothing the map in real space by symmetric kernel
to allow targeting specific regions of the multipole. BiPS is
interesting because it is orientation independent, {\it i.e.}
invariant under rotations of the sky. SI condition implies a null
BiPS, {\it i.e.}  $\kappa_\ell\,=\,0$ for every $\ell>0$, ($
\kappa_\ell\,=\,\kappa_0 \delta_{\ell 0}$). Non-zero components of
BiPS imply break down of statistical isotropy, and this introduces
BiPS as a measure of statistical isotropy. It is worth noting that
although BiPS is quartic in $a_{lm}$, it is designed to detect SI
violation and not non-Gaussianity \cite{us_apjl, us_pascos,us_apj,
us_jgrg}. An un-biased estimator of BiPS is given by \be
\label{estimatork} \tilde\kappa_\ell = \sum_{ll^\prime M}
\left|\biposh{}{\ell M}{ll'}\right|^2 - {\mathfrak B}_\ell\, , \ee
where ${\mathfrak B}_\ell$ is the bias that arises from the SI part of
the map. The bias is given by the angular power spectrum, $C_l$ that
can be estimated from the map (done in the results quoted here), or,
corresponding to the best fit theoretical model (done for WMAP-1
analysis)~\cite{us_apjl,us_apj,haj_sour06,bas06}.

%%%%%%%%%%%%%
\begin{figure}
\begin{center}
\includegraphics[scale=0.25, angle=0]{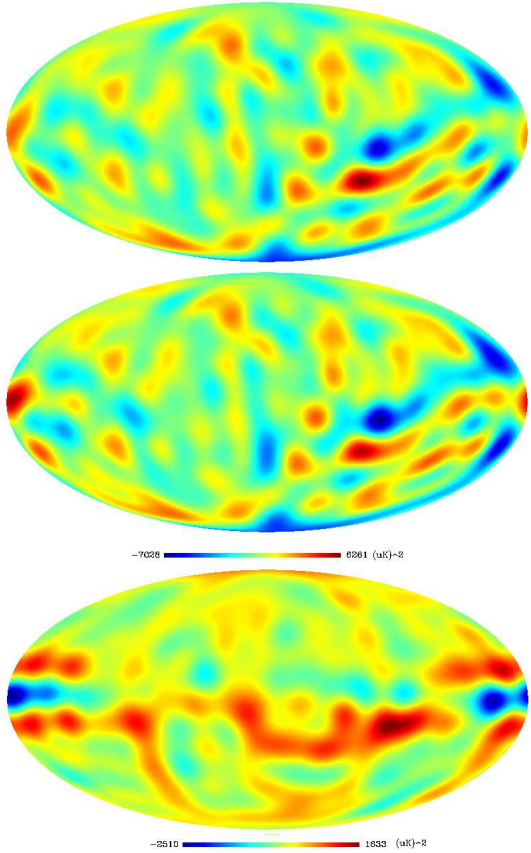}
    \caption{{\bf Top:} A {\it bipolar} map generated from bipolar
    coefficients, $A_{\ell M}$, of 3-year ILC map.  {\bf Middle:}
    bipolar map based on 1-year ILC map and {\bf Bottom:} differences
    between the two maps (note the scales).  The top map (ILC-3) has
    smaller fluctuations comparing to the middle one (ILC-1) except
    for the hot spot near the equator.  Differences between these two
    maps mostly arise from a band around the equator {\it in bipolar
    space}.  Both ILC maps are smoothed by a band pass filter,
    $W^{S}(l_t=2, l_s=10)$. }
  \label{bipolarmap}
\end{center}
\end{figure}
%%%%%%%%%%%%%

\subsection{ Reduced Bipolar Coefficients}
The BipoSH coefficients can be summed over $l$ and $l'$ to reduce the
cosmic variance, \be \label{first} A_{\ell M}=
\sum_{l=0}^{\infty}\sum_{l'=|\ell-l|}^{\ell+l} \biposh{}{\ell
M}{ll'}. \ee 
In any given CMB anisotropy map, $A_{\ell M}$ would fluctuate about
zero. A severe breakdown of statistical isotropy will result in huge
deviations from zero. Reduced bipolar coefficients are not
rotationally invariant, hence they assign direction to the correlation
patterns of a map. An interesting way of visualizing these coefficients is
to make a {\em Bipolar map} from $A_{\ell M}$ \be \Theta(\hat{n}) =
\sum_{\ell=0}^{\infty}\sum_{M=-\ell}^{\ell} A_{\ell M} Y_{\ell M}
(\hat{n}).  \ee The symmetry $A_{\ell M}=(-1)^M A_{\ell -M}^*$ of
reduced bipolar coefficients guarantees reality of $\Theta(\hat{n})$.
The {\em bipolar map} based on bipolar coefficients of ILC-3 is shown
on the top panel of Fig. \ref{bipolarmap}. The map has small
fluctuations except for a pair of hot and cold spots near the
equator. To compare, we have also made a bipolar map of 1-year ILC map
(ILC-1) from bipolar coefficients of ILC-1 (middle panel of
Fig. \ref{bipolarmap}). The difference map (Fig. \ref{bipolarmap}
(bottom)) shows that differences between these two maps mostly arise
from a band around the equator {\it in bipolar space}.  As it is seen
in Fig. \ref{bipolarmap}, the bipolar map of ILC-3 has less
fluctuations comparing to that of ILC-1. This is because almost all of
$A_{\ell M}$'s of ILC-3 are smaller than those of ILC-1 ({\it i.e.}
are closer to zero).
%%%%%%%%%%%%%
\begin{figure}
\includegraphics[scale=0.28, angle=-90]{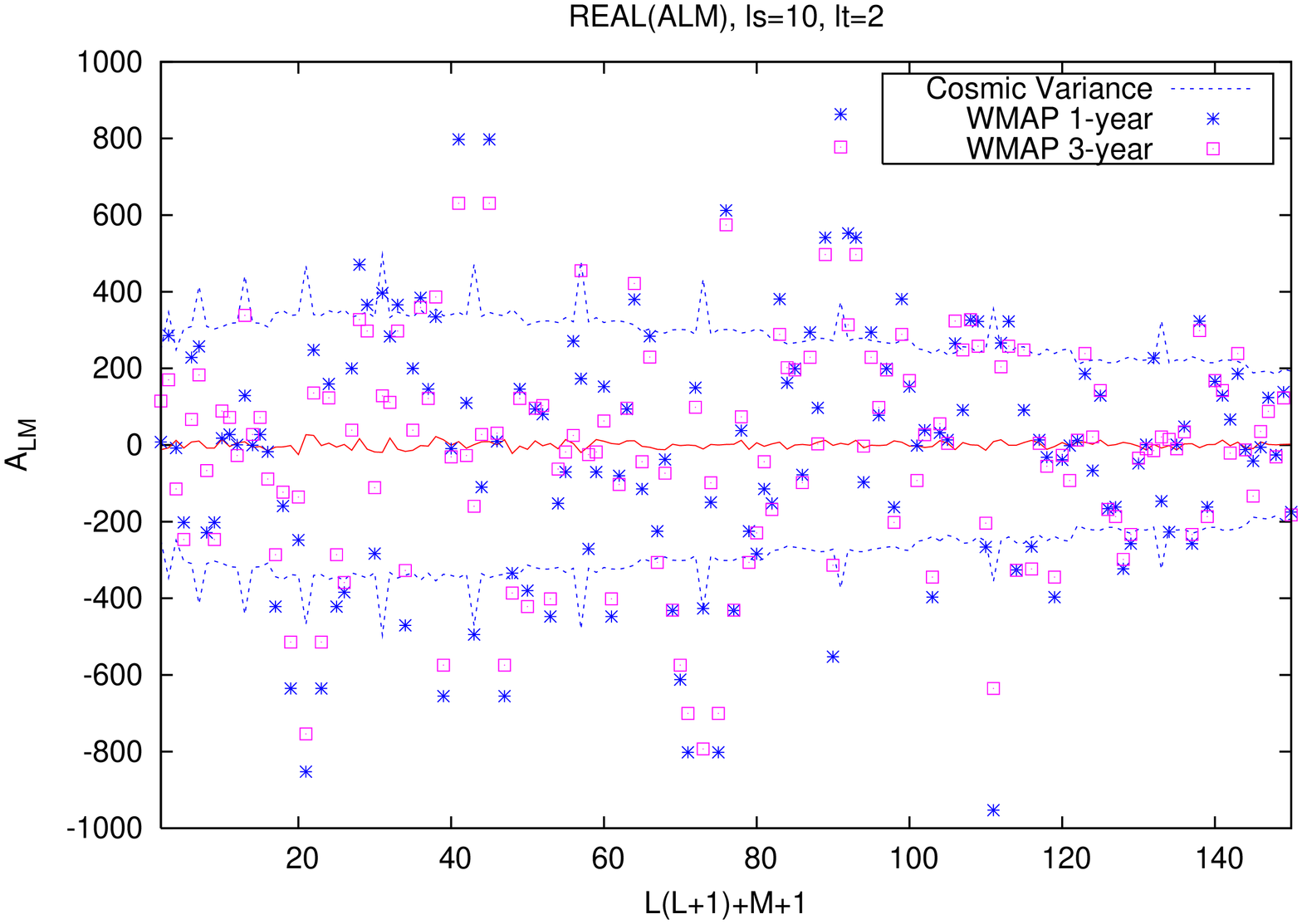}
\includegraphics[scale=0.28, angle=-90]{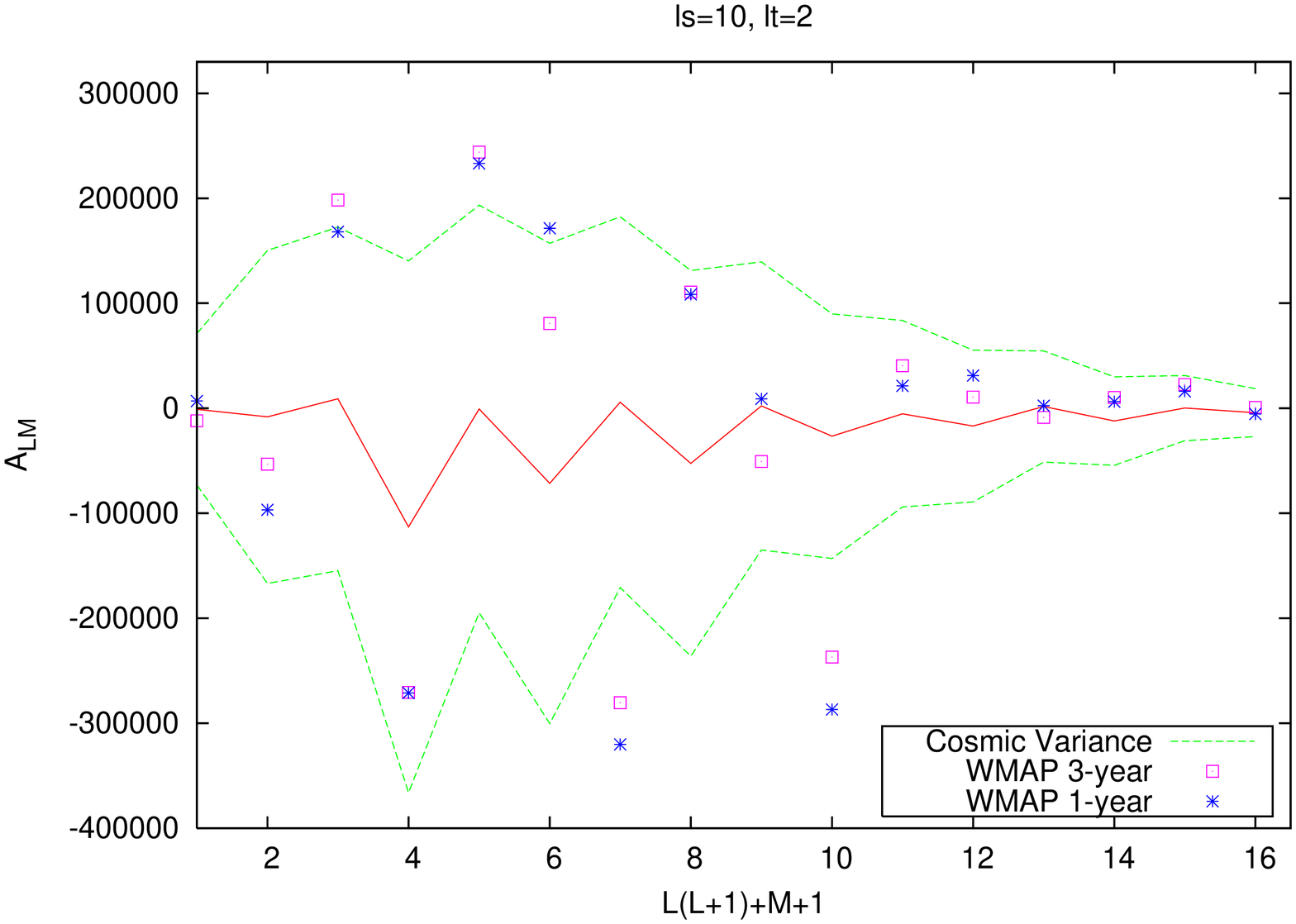}
    \caption{{\em Left:} Real part of $A_{\ell M}$'s of ILC-3 (red
    square points) and ILC-1 (blue stars) for a $W_l^S(10,2)$ filter
    that roughly keeps multipoles between 2 and 15.  $\ell$ and $M$
    indices are combined to a single index $n=\ell(\ell+1)+M+1$.  The
    blue dotted lines define 1-$\sigma$ error bars derived from 1000
    simulations of SI CMB anisotropy maps. Almost all of $A_{\ell
    M}$'s of ILC-3 are smaller than those of ILC-1 which means ILC-3
    is more consistent with statistical isotropy. {\em Right:} The
    BiPS of ILC-3 (red square points) and ILC-1 (blue stars) for a
    $W_l^S(10,2)$ filter is consistent with statistical isotropy The
    green dotted lines define 1-$\sigma$ error bars derived from 1000
    simulations of SI CMB anisotropy maps.  }
  \label{ALM_summed}
\end{figure}

Figure~\ref{ALM_summed} compares $A_{\ell M}$ and $\kappa_\ell$ of
three year WMAP ILC map against an average of 1000 simulations of
statistically isotropic maps with same angular power spectrum. As it
can be seen many spikes presented in $A_{\ell M}$'s of ILC-1 have
either disappeared or reduced in ILC-3 ({\it e.g.} those around $n=20,
40$ and a big spike at $n=111$). More detailed description of analysis
and results are given in ref.~\cite{haj_sour06}.

The rotational properties of $A_{\ell M}$ suggests defining another
rotationally invariant bipolar power spectrum in $D_{\ell}=\sum_{\ell
M}A_{\ell M}A^*_{\ell M}$ -- the same way $a_{lm}$ are combined to
construct the angular power spectrum, $C_l$.  The measurement of {\em
reduced Bipolar Power Spectrum (rBiPS)}, $D_\ell$, on CMB maps is
ongoing. We also defer a comparison of the two orientation insensitive
measures of SI violation $\kappa_\ell$ (BiPS) and $D_\ell$ (rBiPS).

\section{BiPS of CMB polarization maps}
\label{}

One of the firm predictions of this working `standard' cosmological
model is linear polarization pattern ($Q$ and $U$ Stokes parameters)
imprinted on the CMB at the last scattering surface.  A net pattern of
linear polarization is retained due to local quadrupole intensity
anisotropy of the CMB radiation Thomson scattering on the electrons at
at the last scattering surface.

Recently WMAP has provided the first ``full'' sky CMB polarization
maps. The wealth of information in the CMB polarization field will
enable us to determine the cosmological parameters and test and
characterize the initial perturbations and inflationary mechanisms
with great precision. Cosmological polarized microwave radiation in a
simply connected universe is expected to be statistically
isotropic. This is a very important feature which allows us to fully
describe the field by its power spectrum that can have profound
theoretical implications for cosmology. Statistical isotropy (SI) can
now be tested with CMB polarization maps over large sky
fraction. Importance of having statistical tests of departures from SI
for CMB polarization maps lies not only in interesting theoretical
motivations but also in testing the cleaned CMB polarization maps for
residuals from polarized foreground emission.

The coordinate--free description of CMB polarization decomposes the
two kinds of polarization pattern on the sky based on their different
parities.  In the spinor approach, the even parity pattern is called
the $E$--mode and the odd parity pattern the $B$--mode. Hence the CMB
sky maps are characterized by a triplet of  random scalar fields:
$X(\hat{n})\equiv \{T(\hat{n})$, $E(\hat{n})$, $B(\hat{n})\}$.
Statistical properties of each of these fields can be characterized by
$N$-point correlation functions, $\langle X(\hat{n}_1)
X(\hat{n}_2)\cdots X(\hat{n}_n)\rangle$\footnote{For cut-sky,
$E(\hat{n})$ and $B(\hat{n})$ mode decomposition is not unique
\cite{Lewis:2002,Brown:2004}. But since mixing is linear there always
exist two linearly independent modes. It is possible to formulate the
SI of these linear independent modes.}.
\begin{figure}
\begin{center}
\includegraphics[scale=0.35, angle=0]{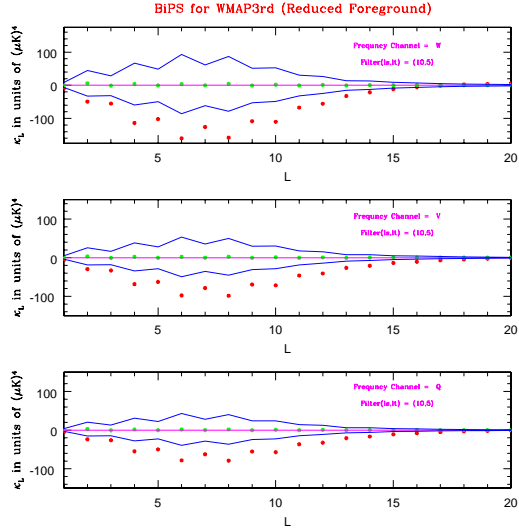}
\caption{BiPS for polarization maps(nside = 512) with polarization mask 
  using band-pass filter $W^{S}_{l}(l_s=10,l_t=5)$. Red dots show the
  BiPS for WMAP3rd foreground cleaned polarization map with
  polarization mask after bias subtraction, green dots show the BiPS
  for 100 SI maps with polarization mask after bias subtraction and
  Blue lines show the $1-\sigma$ of the cosmic variance of BiPS for
  100 SI maps with polarization mask.}
\label{bips_wmap3pol}
\end{center}
\end{figure}

Gaussian CMB sky is completely described by two-point correlation
functions of $X(\hat{n})$, or equivalently, the corresponding
spherical harmonic coefficients \be C^{XX'}(\hat{n},\hat{n'})\,=\,
\langle X(\hat{n}) X'(\hat{n}') \rangle' \langle
a^X_{lm}a^{X'*}_{l^\prime m^\prime}\rangle.  \ee It is possible to
define generalized Bipolar Spherical Harmonic (BipoSH) coefficients,
$\biposh{XX'}{\ell M}{l_1 l_2}$, that are linear combinations of
off-diagonal elements of the $\langle a^X_{lm}a^{X'*}_{l^\prime
m^\prime}\rangle$ covariance matrix, \be
\label{ALMvsalm_gen} \biposh{XX'}{\ell M}{l_1 l_2} \,=\, \sum_{m_1m_2} 
\langle a^X_{l_1m_1}a^{*X'}_{l_2 m_2}\rangle (-1)^{m_2} {\mathcal
  C}^{\ell M}_{l_1m_1l_2 -m_2} \ee where ${\mathcal
  C}_{l_1m_1l_2m_2}^{\ell M}$ are Clebsch-Gordan coefficients.
  Incomplete sky coverage induces a contamination of E-mode of
  polarization by its B-mode and vice-versa. Then the modified
  temperature and polarization fields is related to their actual
  values of full sky coverage by a window matrix
  \cite{Lewis:2002,Brown:2004} whose elements are basically window
  functions for temperature and polarization in harmonic space. It can
  be shown that the estimated BipoSH coefficients are in fact linear
  combinations of that for full-sky CMB maps,  \be
  \tildebiposhmat{}{\ell M}{l l'} = \sum_{\ell' M' l_1 l_2} {\bf
  N}^{\ell M l l'} _{\ell' M' l_1 l_2}\,\, \biposhmat{}{\ell M}{l_1
  l_2}\,. \label{biposhmat}\ee Here bold-faced $
  \tildebiposhmat{}{\ell M}{l l'}$ and $\biposhmat{}{\ell M}{l_1 l_2}$
  are the column matrices corresponding to estimated and true BipoSH
  coefficients respectively, for the auto and cross-correlations $(TT,
  TE, TB, ET, EE, EB, BT, BE, BB)$ of temperature anisotropy and
  polarization. The elements of the matrix $ {\bf N}^{\ell M l
  l'}_{\ell' M' l_1 l_2}$ depend on Clebsch- Gordan coefficients and
  window functions in harmonic space. Hence, the true BipoSH
  coefficients can be estimated from the pseudo-BipoSH coefficients by
  inverting the above equation.

Figure~\ref{bips_wmap3pol} shows BIPS for the foreground cleaned
polarization maps recently released by WMAP filtered to retain power
multipole between $l\sim 5$ to $15$.  The BiPS for CMB E-polarization
maps at $V$ and $W$ from WMAP are compared to the mean and 1-$\sigma$
deviations of BiPS measurements on $100$ simulated maps The
appropriate galactic mask for polarization maps suggested by WMAP was
employed for all maps.  These {\em preliminary} results indicate strong
violation of statistical isotropy in the polarization. However, the
dependence on frequency suggests a non cosmological origin in
foreground residuals in parts of the sky beyond galactic
mask. Interestingly enough,c onsistent with conclusions of the WMAP
polarization analysis, the BiPS analysis also indicates that the $V$
band is `cleaner' than the $W$ band.

\section{Discussion}
\label{sec:discussion}

The Bipolar representation of the statitical correlations underlying
the CMB anisotropy and polarization allows a unprejudiced
evaluation of statistical isotropy on largest observable cosmic
scales. The null result of search for departure from statistical
isotropy in the WMAP data provides strong evidence for cosmological
principle and constrains non-trivial cosmic topology and ultra-large
scale structure.  It can potentially be used to constrain any other
physics that violate global isotropy on cosmic scales.  The null
result has implications for the observation and data analysis
techniques used to create the CMB anisotropy maps.  Observational
artifacts such as non-circular beam, inhomogeneous noise correlation,
residual striping patterns, and residuals from foregrounds are
potential sources of SI breakdown.


\begin{thebibliography}{99}
\frenchspacing

\bibitem{bps98} J. R. Bond, D. Pogosyan \& T. Souradeep, Class. Quant.
  Grav. {\bf 15}, 2671 (1998).

\bibitem{bps00a} J. R. Bond, D. Pogosyan \& T. Souradeep, Phys.  Rev. {\bf D
  62},043005 (2000).

\bibitem{bps00b} J. R. Bond, D. Pogosyan \& T. Souradeep, Phys.
Rev. {\bf D 62},043006 (2000).

\bibitem{hin_wmap03} G. Hinshaw, Astrophys. J. Suppl. {\bf 148}, 135 (2003) .

\bibitem{kogut_wmap03} A. Kogut et al., Astrophys. J. Suppl.,{\bf
148}, 161 (2003).

\bibitem{sper_wmap03} D.  Spergel et al., Astrophys. J. Suppl., {\bf
148}, 175, (2003).

\bibitem{page_wmap03} L. Page et al., Astrophys. J. Suppl. {\bf 148},
233 (2003).

\bibitem{peiris_wmap03} H.V. Peiris  et al.,  Astrophys. J. Suppl. {\bf 148}, 213 (2003).

\bibitem{sper_wmap06} D. Spergel et al., {\it preprint}, [arXiv:
astro-ph/0603449).


\bibitem{hin_wmap06} G. Hinshaw  et al., {\it preprint} , [arXiv:astro-ph/0603451].

\bibitem{pag_wmap06} L. Page et al., {\it preprint}, [arXiv:astro-ph/0603450].

\bibitem[{Eriksen et al.} (2004a)]{erik04a}
  H.~K.~Eriksen, F.~K.~Hansen, A.~J.~Banday, K.~M.~Gorski and P.~B.~Lilje,
  %``Asymmetries in the CMB anisotropy field,''
  Astrophys.\ J.\  {\bf 605}, 14 (2004)
  [Erratum-ibid.\  {\bf 609}, 1198 (2004)]
  % [arXiv:astro-ph/0307507].

%\cite{Copi:2003kt}
\bibitem[{Copi et al.} (2004)]{Copi:2003kt}
  C.~J.~Copi, D.~Huterer and G.~D.~Starkman,
  %``Multipole Vectors--a new representation of the CMB sky and evidence for
  %statistical anisotropy or non-Gaussianity at 2<=l<=8,''
  Phys.\ Rev.\ D {\bf 70}, 043515 (2004)
  %[arXiv:astro-ph/0310511].


%\cite{Schwarz:2004gk}
\bibitem[{Schwarz et al.} (2004)]{Schwarz:2004gk}
  D.~J.~Schwarz, G.~D.~Starkman, D.~Huterer and C.~J.~Copi,
  %``Is the low-l microwave background cosmic?,''
  Phys.\ Rev.\ Lett.\  {\bf 93}, 221301 (2004)
  % [arXiv:astro-ph/0403353].

  %\cite{Hansen:2004vq}
\bibitem[{Hansen et al.} (2004)]{Hansen:2004vq}
  F.~K.~Hansen, A.~J.~Banday and K.~M.~Gorski,
  %``Testing the cosmological principle of isotropy: local power spectrum
  %estimates of the WMAP data,''
  arXiv:astro-ph/0404206.

\bibitem[{de Oliveira-Costa et al.} (2003)]{angelwmap} A.~de Oliveira-Costa,
M.~Tegmark, M.~Zaldarriaga, \& A.~Hamilton, 2004, Phys. Rev.{\bf D69}, 063516.

%\cite{Land:2004bs}
\bibitem[{Land \& Magueijo} (2005a)]{Land:2004bs}
  K.~Land and J.~Magueijo,
  %``Cubic anomalies in WMAP,''
  Mon.\ Not.\ Roy.\ Astron.\ Soc.\  {\bf 357}, 994 (2005)
  % [arXiv:astro-ph/0405519].

%\cite{Land:2005ad}
\bibitem[{Land \& Magueijo} (2005b)]{Land:2005ad}
  K.~Land and J.~Magueijo,
  %``The axis of evil,''
  Phys.\ Rev.\ Lett.\  {\bf 95}, 071301 (2005)
  %[arXiv:astro-ph/0502237].
 %#
%\cite{Land:2005dq}
\bibitem[{Land \& Magueijo} (2005c)]{Land:2005dq}
  K.~Land and J.~Magueijo,
  %``The Multipole Vectors of WMAP, and their frames and invariants,''
  Mon.\ Not.\ Roy.\ Astron.\ Soc.\  {\bf 362}, 838 (2005)
  %[arXiv:astro-ph/0502574].

%\cite{Land:2005jq}
\bibitem[{Land \& Magueijo} (2005d)]{Land:2005jq}
  K.~Land and J.~Magueijo,
  %``Is the Universe odd?,''
  Phys.\ Rev.\ D {\bf 72}, 101302 (2005)
  %[arXiv:astro-ph/0507289].

 %\cite{Land:2005cg}
\bibitem[{Land \& Magueijo} (2005e)]{Land:2005cg}
  K.~Land and J.~Magueijo, Mon.Not.Roy.Astron.Soc. {\bf 367}  1714 (2006)
  %``Template fitting and the large-angle CMB anomalies,''
  %arXiv:astro-ph/0509752.

%\cite{Bielewicz:2004en}
\bibitem[{Bielewicz et al.} (2004)]{Bielewicz:2004en}
  P.~Bielewicz, K.~M.~Gorski and A.~J.~Banday,
  %``Low order multipole maps of CMB anisotropy derived from WMAP,''
  Mon.\ Not.\ Roy.\ Astron.\ Soc.\  {\bf 355}, 1283 (2004)
  %[arXiv:astro-ph/0405007].

%\cite{Bielewicz:2005zu}
\bibitem[{Bielewicz et al.} (2005)]{Bielewicz:2005zu}
  P.~Bielewicz, H.~K.~Eriksen, A.~J.~Banday, K.~M.~Gorski and P.~B.~Lilje,
  %``Multipole vector anomalies in the first-year WMAP data: a cut-sky
  %analysis,''
  Astrophys.\ J.\  {\bf 635}, 750 (2005)
  %[arXiv:astro-ph/0507186].

%\cite{Copi:2005ff}
\bibitem[{Copi et al.} (2006)]{Copi:2005ff}
  C.~J.~Copi, D.~Huterer, D.~J.~Schwarz and G.~D.~Starkman,
  %``On the large-angle anomalies of the microwave sky,''
  Mon.\ Not.\ Roy.\ Astron.\ Soc.\  {\bf 367}, 79 (2006)
  %[arXiv:astro-ph/0508047].


%\cite{Copi:2006tu}
\bibitem[{Copi et al.} (2006b)]{Copi:2006tu}
  C.~Copi, D.~Huterer, D.~Schwarz and G.~Starkman, {\it preprint}
 %   ``The Uncorrelated Universe: Statistical Anisotropy and the Vanishing Angular
 %   Correlation Function in WMAP Years 1-3,''
  %
  [arXiv:astro-ph/0605135].


%\cite{Naselsky:2004gm}
\bibitem[{Naselsky et al.} (2004)]{Naselsky:2004gm}
  P.~D.~Naselsky, L.~Y.~Chiang, P.~Olesen and O.~V.~Verkhodanov,
  %``Primordial magnetic field and non-Gaussianity of the 1-year Wilkinson
  %Microwave Anisotropy Probe (WMAP) data,''
  Astrophys.\ J.\  {\bf 615}, 45 (2004)
  %[arXiv:astro-ph/0405181].

%\cite{Prunet:2004zy}
\bibitem[{Prunet et al.} (2005)]{Prunet:2004zy}
  S.~Prunet, J.~P.~Uzan, F.~Bernardeau and T.~Brunier,
  %``Constraints on mode couplings and modulation of the CMB with WMAP data,''
  Phys.\ Rev.\ D {\bf 71}, 083508 (2005)
  %[arXiv:astro-ph/0406364].

%\cite{Gluck:2005td}
\bibitem[{Gluck \& Pisano}(2005)]{Gluck:2005td}
  M.~Gluck and C.~Pisano,
  %``Alternative Multipole Vectors for the CMB Temperature Fluctuations,''
  [arXiv:astro-ph/0503442].

%\cite{Stannard:2004yp}
\bibitem[{Stannard \& Coles}(2005)]{Stannard:2004yp}
  A.~Stannard and P.~Coles,
  %``Random-Walk Statistics and the Spherical Harmonic Representation of CMB
  %Maps,''
  Mon.\ Not.\ Roy.\ Astron.\ Soc.\  {\bf 364}, 929 (2005)
  %[arXiv:astro-ph/0410633].

 %\cite{Bernui:2005pz}
\bibitem[{Bernui et al.}(2005)]{Bernui:2005pz}
  A.~Bernui, B.~Mota, M.~J.~Reboucas and R.~Tavakol,
  %``Mapping large-scale anisotropy in the WMAP data,''
  {\it preprint} [arXiv:astro-ph/0511666].

%\cite{Bernui:2006ft}
\bibitem[{Bernui et al.}(2006)]{Bernui:2006ft}
  A.~Bernui, T.~Villela, C.~A.~Wuensche, R.~Leonardi and I.~Ferreira,
  %``On the CMB large-scales angular correlations,''
  {\it preprint} [arXiv:astro-ph/0601593].


%\cite{Freeman:2005nx}
\bibitem[{Freeman et al.}(2006)]{Freeman:2005nx}
  P.~E.~Freeman, C.~R.~Genovese, C.~J.~Miller, R.~C.~Nichol and L.~Wasserman,
  %``Examining the Effect of the Map-Making Algorithm on Observed Power
  %Asymmetry in WMAP Data,''
  Astrophys.\ J.\  {\bf 638}, 1 (2006)
  % [arXiv:astro-ph/0510406].

 %\cite{Chen:2005ev}
\bibitem[{Chen \& Szapudi} (2005)]{Chen:2005ev}
  G.~Chen and I.~Szapudi,
  %``Measuring the Three Point Correlation Function of the Cosmic Microwave
  %Background,''
  Astrophys.\ J.\  {\bf 635}, 743 (2005)
  %[arXiv:astro-ph/0508316].

%\cite{Wiaux:2006zh}
\bibitem{Wiaux:2006zh}
  Y.~Wiaux, P.~Vielva, E.~Martinez-Gonzalez and P.~Vandergheynst,
   ``Global universe anisotropy probed by the alignment of structures in the
  %cosmic microwave background,''
  Phys.\ Rev.\ Lett.\  {\bf 96}, 151303 (2006)
  % [arXiv:astro-ph/0603367].



\bibitem{us_apjl} A. Hajian \& T. Souradeep, Astrophys. J. Lett. {\bf
 597}, L5 (2003).


\bibitem{us_pascos} T. Souradeep and A. Hajian, Pramana, {\bf 62},
793. (2004).

\bibitem{us_prl} A. Hajian \& T. Souradeep, {\it preprint}
[arXiv:astro-ph/0301590].

\bibitem[{Ghosh et al.} (2006)]{ghos06}
  T.~Ghosh, A.~Hajian and T.~Souradeep,{\it preprint} [arXiv:astro-ph/0604279].
  %``Unveiling Hidden Patterns in CMB Anisotropy Maps,''
  % arXiv:astro-ph/0604279.
  %%CITATION = ASTRO-PH 0604279;%%


\bibitem{sah06}  R. Saha, P. Jain \& T. Souradeep, Astrophys. J. Lett.,
{\bf 645}, L89-L92, (2006).


\bibitem{us_apjl2} A. Hajian \&  T. Souradeep \& N. Cornish,
  Astrophys. J. Lett. {\bf 618}, L63 (2005). 

\bibitem{us_apj} A. Hajian \& T. Souradeep, {\it preprint}
[arXiv:astro-ph/0501001].

\bibitem{us_jgrg} T. Souradeep and A. Hajian, Proceedings of JGRG-14, (2004)

\bibitem{haj_sour06} A. Hajian \& T. Souradeep, {\it preprint}
[arXiv:astro-ph/0607153].

\bibitem{bas06} S. Basak, A. Hajian \&  T. Souradeep, Phys. Rev. {\bf D 74},
021301(R) (2006).

\bibitem{Var}  D. A. Varshalovich, A. N Moskalev,
 V. K. Khersonskii, {\it Quantum Theory of Angular
  Momentum} (World Scientific, 1988).

\bibitem{Lewis:2002}
A. Lewis, A. Challinor \&  N.  Turok, Phys. Rev. {\bf D 65}, 023505 (2002).

\bibitem{Brown:2004} 
M. L. Brown, P. G. Castro, A. N. Taylor,
Mon.Not.Roy.Astron.Soc. {\bf 360}, 1262, (2005).


\end{thebibliography}
\end{document}